\documentclass[amsmath,prb,twocolumn]{revtex4-1}
\usepackage{amsmath}
\usepackage{mathrsfs}
\usepackage{amsfonts}
\usepackage{graphicx}
\usepackage{amssymb}
\usepackage{epstopdf}
\usepackage{color}
\definecolor{rojo}{RGB}{255,40,0}

\begin{document}

\title{Sound field radiated by a rod}

\newcommand{\TGb}[1]{{\color{blue} [TG] {#1}}}
\newcommand{\TGr}[1]{{\color{red} [TG] {#1}}}

\author{L.~A. Razo-L\'opez}
\email{lalbertorazo@gmail.com}
\affiliation{Departamento de F\'isica, Universidad de Guadalajara,
	Blvd. Marcelino Garc\'ia Barragan y Calzada Ol\'impica, C.P. 44840, Guadalajara, Jalisco, M\'exico and\\
Universit\`e C\^ote d'Azur, Institut de Physique de Nice, CNRS, Nice, 06100, France.}

\author{V. Dom\'inguez-Rocha}
\email{vdr@xanum.uam.mx}
\affiliation{Instituto de Ciencias F\'isicas, Universidad Nacional Aut\'onoma de M\'exico, Apartado Postal 48-3, 62210 Cuernavaca, Mor., Mexico and\\
	Departamento de F\'isica, Universidad Aut\'onoma Metropolitana-Iztapalapa,
	Apartado Postal 55-534, 09340 Ciudad de M\'exico, Mexico.}

\author{John A. Franco-Villafa\~ne}
\email{jofravil@ifisica.uaslp.mx}
\affiliation{CONACYT - Instituto de F\'isica, Universidad Aut\'onoma de San Luis Potos\'i, 78290 San Luis Potos\'i, M\'exico}

\author{T. Gorin}
\email{thomas.gorin@cucei.udg.mx}
\affiliation{Departamento de F\'isica, Universidad de Guadalajara,
	Blvd. Marcelino Garc\'ia Barragan y Calzada Ol\'impica, C.P. 44840, Guadalajara, Jalisco, M\'exico}

\author{R.~A. M\'endez-S\'anchez}
\email{mendez@icf.unam.mx}
\affiliation{Instituto de Ciencias F\'isicas, Universidad Nacional Aut\'onoma de M\'exico, Apartado Postal 48-3, 62210 Cuernavaca, Mor., Mexico}

\begin{abstract}
We study the acoustical intensity field radiated by a thin cylindrical rod vibrating in its lowest compressional mode. Due to the cylindrical symmetry, the emitted field is measured in a radial plane of the rod which is sufficient to reconstruct the full three-dimensional field. Starting from the one-dimensional approximation of the excited compressional mode, we develop a simplified theoretical wave equation which allows for a semi-analytical solution for the emitted wave field. The agreement between the experimental results and the semi-analytical solution is eloquent.
\end{abstract}

\maketitle

\section{Introduction}

Even when the significant progress of acoustics was developed in the 60's, some exciting features remain unexplored for scientists and engineers. A good example is the acoustical field distribution coming from a freely vibrating rod, where the studies are scarce, due to the difficulty to excite a solid with a well defined frequency as well as with a non-contact perturbation keeping a reasonable signal-to-noise ratio.

The radiated field of a rod can be used to study its resonant frequencies~\cite{Anderson}, and it has been directly measured in water using the standard technique of exciting the solid with a piezoelectric transducer~\cite{Jarosz, Gook, Malladi}. Another common technique to excite a solid is the use of a shaker instead of a piezoelectric transducer~\cite{Wu, Blake}. Despite the efficiency of those methods to excite a medium, they cannot be used to analyze the radiated acoustic field of a freely vibrating rod. That is because the physical contact between the exciter and the rod induces undesired higher vibrating modes that disturb the wave field of interest~\cite{Morales}.

A non-contact technique to excite structures consists of the use of a collimated air pulses controlled by valves~\cite{Farshidi}. However, due to strong air perturbations, it does not allow us to measure the sound field properties adequately. A more suitable technique uses electromagnetic acoustic transducers (EMATs)~\cite{Hefner}. This technique allows to excite and detect the vibrations of a solid with extraordinary accuracy due to the high selectivity in the frequency as well as its noise-reduction capacities. 

From the theoretical point of view, the radiated acoustic field of a vibrating rod was studied extensively a long time ago by Williams et al.~\cite{Williams}. But it can also be considered by more straightforward models using a finite sum of spherical harmonics~\cite{Morfey}, or the Green's functions method~\cite{New}, among others.

In this paper, the experimental stationary acoustical field produced by a freely vibrating rod is studied. The rod has finite length and is vibrating in its first compressional mode. Due to the nature of this mode, the amplitude of the harmonic displacement is maximal at the ends of the rod. This means that the biggest interaction with the surrounded media is occurring there. In order to describe the acoustical wave field, a semi-analytical model is developed. Finally, the experimental results are compared with the theoretical prediction.

\section{Elastic modes of the rod}

\begin{figure}[h!]
	\begin{center}
		\includegraphics[width=0.37\textwidth]{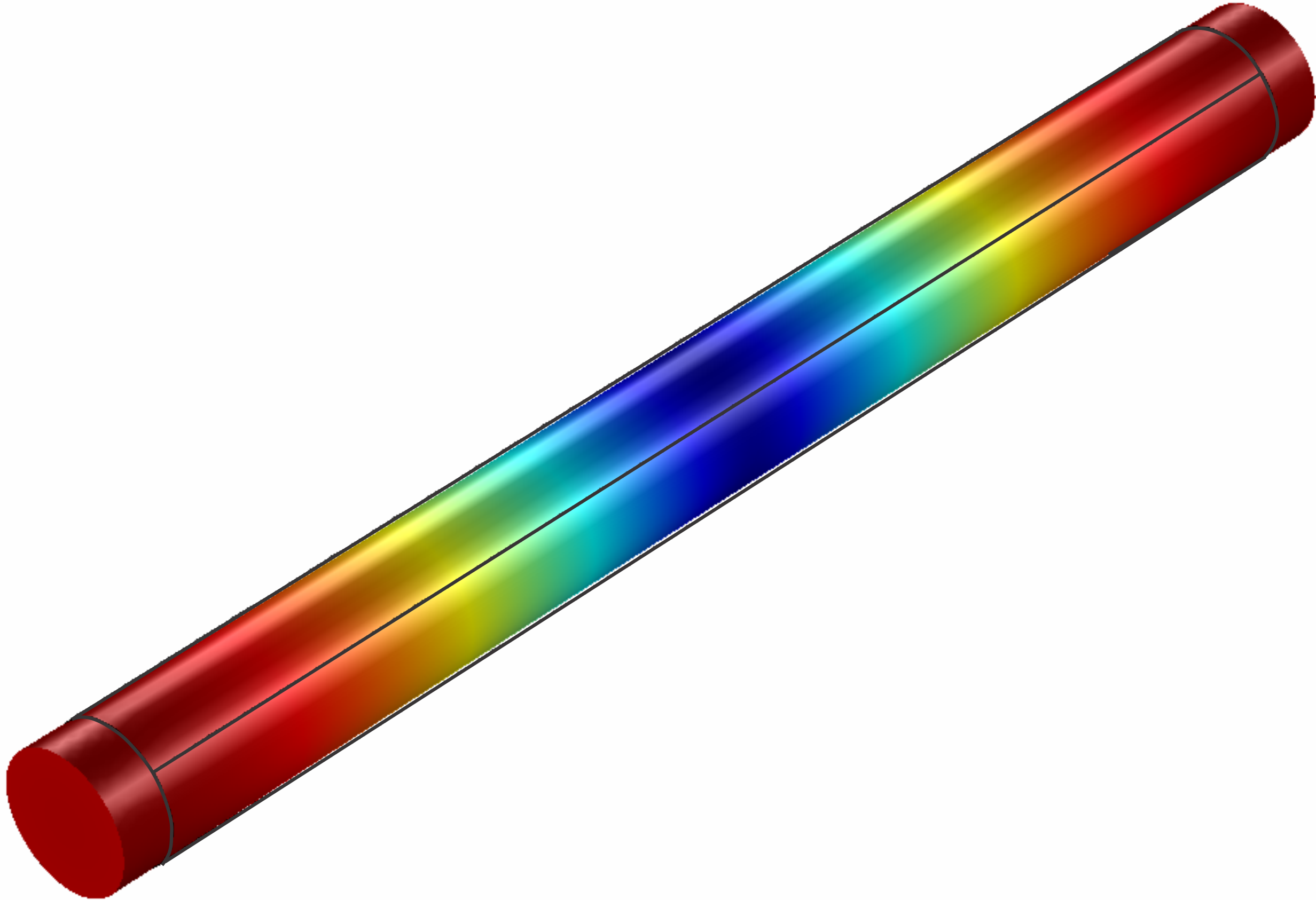}
		\includegraphics[width=0.1\textwidth]{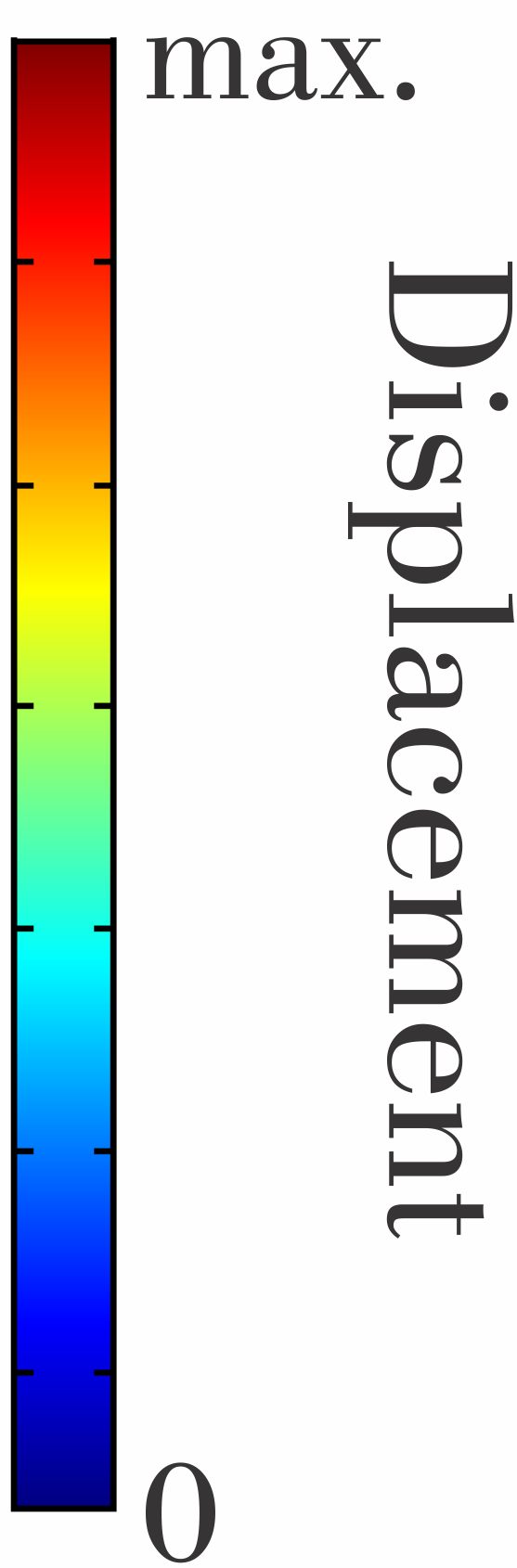}
		\caption{\label{fig:fig1}{(Colour online) COMSOL simulation of a cylindrical rod vibrating in its first compressional normal mode. The relative surface displacement is shown from blue to red. The number and position of nodes are predicted by Eq.~(\ref{eq:nodes}). This normal mode presents a node at the center of the rod and two maxima at the axial ends. The black wires represent the equilibrium state of the rod.}}
	\end{center}
\end{figure}

As a quasi-onedimensional elastic system, where the length $L$ is much larger than the diameter $D$, the rod has eigenmodes, which can be classified into compressional, torsional and flexural modes~\cite{Graff, Morales2002, DominguezRocha, Franco}. The compressional eigenfrequencies are given by
\begin{equation}
\label{eq:freq}
f_\nu = \frac{\nu}{2L}\; \sqrt{\frac{E}{\rho}} \; , \qquad
\nu = 1,2,3,\ldots\; ,
\end{equation}
where $E$ is the Young modulus, $\rho$ is the density, and $\nu$ is the mode
number. For a given mode number $\nu$, the nodes $\xi_\eta$ of the wave
function which describes the local displacement of volume elements in the rod,
are located at
\begin{equation}
\label{eq:nodes}
\xi_\eta = \frac{\eta\, L}{\nu +1} \; , \qquad \eta = 1,2,\ldots, \nu\; .
\end{equation}
Specifically, for our experiment we used an aluminium rod of 1m length and 0.0127m diameter. Hence, according to Eq.~(\ref{eq:freq}), the first compressional mode has the frequency of $f_{1\textrm{C}} = 2.5\, {\rm kHz}$, and according to Eq.~(\ref{eq:nodes}), this mode has one node which lies exactly in the middle of the rod at $\xi_1 = L/2$. Fig.~\ref{fig:fig1} shows the displacement for the first compressional mode of a free boundary rod which presents a node in its geometrical center. 

In vacuum, the dynamical oscillation produces mainly a shrink and stretch process along the axial axis of the rod. Now, let us place this oscillating system into the air. As can be anticipated, the strongest interaction between the air and the rod takes place at both ends of the rod. From this interaction, we neglect the effect of the air on the oscillating rod and focus on the acoustical wave field generated by the oscillating ends of the rod. To that end, we observe that the system has an axial symmetry with respect to the symmetry axis of the rod. In addition, we note that we may neglect reflections of the provided acoustical wave field at other boundaries at sufficiently long times. Finally, we neglect the much smaller radial displacement of the boundary of the open cylinder (that is the cylinder boundary without its ends). Thus, to measure the acoustical wave field of the rod it is sufficient to measure the wave field in a plane that contains the symmetry axis of the rod.

\section{Experimental measurement of the acoustical wave field}

\begin{figure}[h!]
	\begin{center}
		\includegraphics[width=0.41\textwidth]{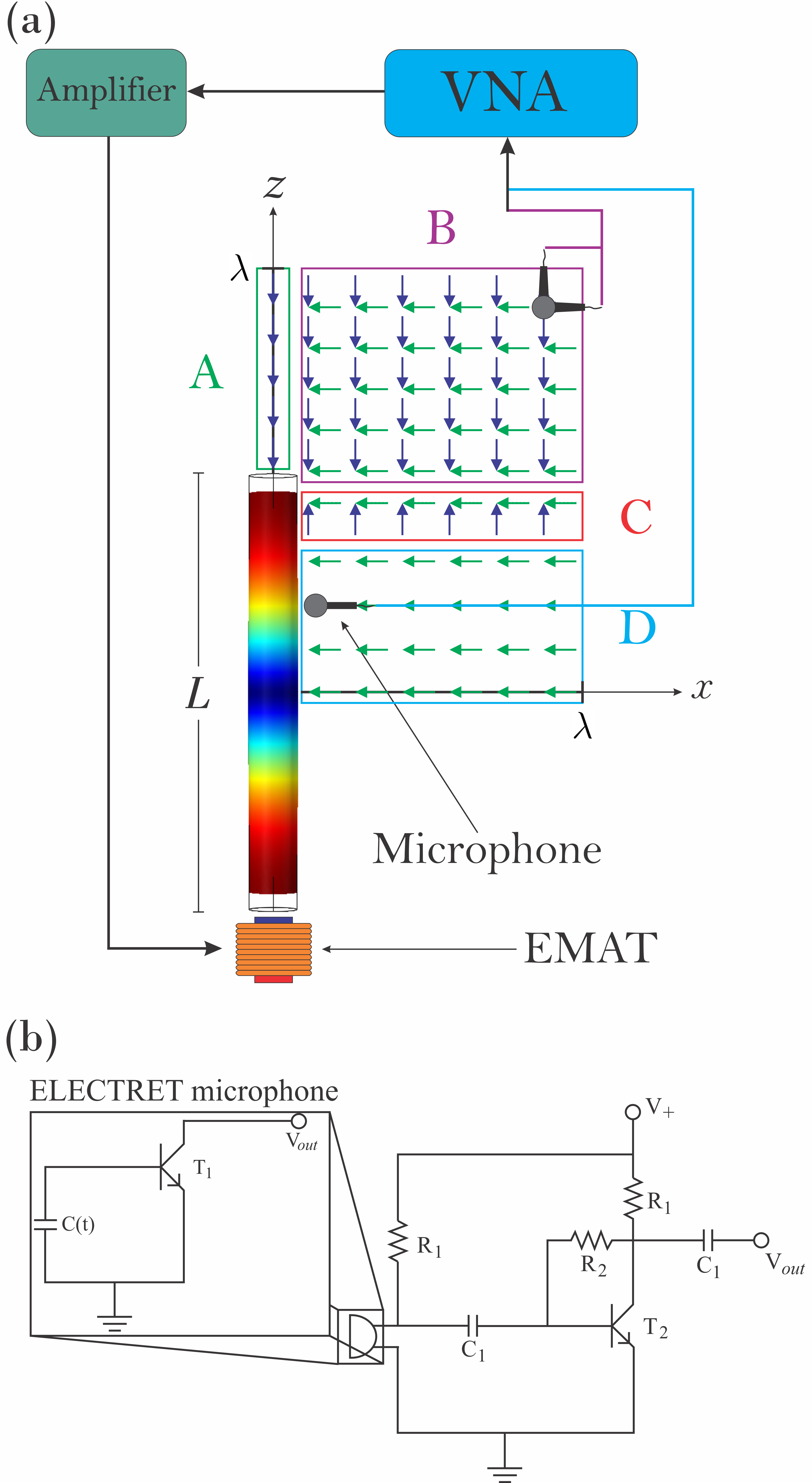}
		\caption{\label{fig:fig2}{(Colour online) (a) Experimental setup to measure the spatial distribution of the acoustic field distribution of a compressional mode of a rod. The displacement field scale of the rod is the same as in \ref{fig:fig1}. One of the rod edges is under the action of a harmonic force, while the opposite one remains free. The sound is measured with a microphone at different positions, of the four regions A, B, C and D. The arrows indicate how the microphone is placed to detect the sound radiation at different space positions in each region. The configuration of the microphone shows two measuring positions in regions B and D. (b) The circuit diagram used to feed the ELECTRET microphone. The resistances and capacitances values are: $R_1=10$~k$\Omega$, $R_2=100$~k$\Omega$, and $C_1=0$.1~$\mu$F. The transistor models are 2SK596 for $T_1$, and 2N3904 for $T_2$. It is also shown the circuit diagram of the ELECTRET microphone where the pressure sensible capacitor $C(t)$ is located.}}
	\end{center}
\end{figure}

The experimental scheme shown in Fig.~\ref{fig:fig2}(a) is used to measure the acoustical wave field emitted by a cylindrycal aluminum rod which is vibrating in its first compressional mode. To generate the compressional resonance, a vector network analyzer (VNA, Anritsu MS-4630B) is used. The VNA produces a sinusoidal signal of frequency $f$, which is amplified by a Cerwin-Vega amplifier (CV-900). Then, the signal is sent to an exciter (EMAT) that is placed very close to the rod. 

The EMAT is basically controlled by the VNA in a small frequency window tuned around 2.5~kHz. A further advantage of EMATs is that they work up to several tens of kHz making them excellent for studying the resonance spectra of a body. The rod is excited without physical contact using an EMAT, which is highly selective between different modes~\cite{Morales, Morales2002}. Also, EMAT's can be used to excite a system in different ways depending on the magnetic properties of the material~\cite{Rossing1992, Russell2000, Russell2013}.

The vibration of the rod interacts with the surrounded air producing an audible signal (acoustic field) that is measured by an ELECTRET microphone. Those microphones contain a transistor inside which acts as an amplifier. In order to operate these devices correctly, a small power source is added to feed the transistor. The diagram of the external amplifier and the power source is shown in Fig.~\ref{fig:fig2}(b).

\begin{figure}[h!]
	\begin{center}
		\includegraphics[width=0.48\textwidth]{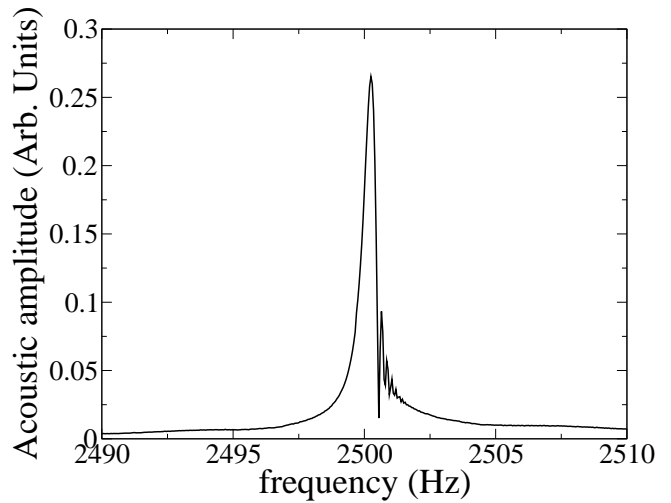}
		\caption{\label{fig:fig3}{The experimental acoustical amplitude measurement by a microphone in region A. The acoustical wave field is generated by a cylindrical rod vibrating in its first compressional mode. Each measurement presents a maximum at the resonance frequency which does not depend on the microphone configuration.}}
	\end{center}
\end{figure}

The rod is excited in the same frequency window many times while the position and orientation of the microphone are changed to map the area of interest. Each measurement taken by the microphone is sent back to the VNA which allows us to obtain the amplitude of the pressure variation. Since one resonant frequency of the rod lies inside the chosen frequency window, we can discriminate the resonance amplitude as the highest point in the measurement. Figure~\ref{fig:fig3} shows a common resonance measurement taken by aligning the center of the microphone with the axial axis of the rod (region A) by keeping a distance of 1cm between them. The actual amplitude value is related to the spatial microphone configuration and assigned to the specific coordinate where the microphone was placed. A detailed explanation about how each measurement was taken and the data analysis is included in the next subsection. We have to mention that we are interested in the intensity of the acoustic wave field that can be gotten by means of the amplitude measured by the VNA.

Another fundamental detail to take into account in acoustic studies is the reflection of waves in the walls of the laboratory. To avoid them, we used polyurethane conventional acoustic sponge, like the one used in music recording studios. In order to reduce a direct reflection, we cover the sponge with low-density polyethylene foam. Finally, the insulating set (polyurethane plus polyethylene) is placed in a wedge shape in order to promote the reflections of the acoustic waves towards their vertex. Four of these wedges are placed around the system and two flat insulating sets are situated, one on the ceiling and another on the work table.

\subsection{Description of the measurement of the sound field}

Given the cylindrical symmetry of the rod, we restrict the measurement area to four regions (see Fig.~\ref{fig:fig2}(a)). Region A is measured along the axial axis from the rod edge and has a length of 0.14~m; while region D is perpendicular to the axial axis from the middle of the rod and form a rectangle whose sides are 0.14~m width and 0.45~m length. While in regions A and D measurements are taken by pointing the microphone towards the rod, in regions B and C two measurements are taken at each point. One measurement is taken by placing the microphone parallel to the axial axis of the rod while the other one is taken by chosing a perpendicular configuration. Region B is a square of 0.14~m side length and region C is a rectangle whose sides are 0.14~m length and 0.05~m width.

The closest measurements in all regions are approximately 1~mm away from the rod. For the measurements of region A, the center of the microphone is aligned with the axial axis of the rod and each measurement has a distance of 1~cm from the contiguous ones. Region B is divided into a square mesh of 1~cm length between adjacent points. Similarly, region C is divided into a rectangular mesh of 1~cm length between adjacent points. Finally, region D is divided into a rectangular mesh of 5~cm by 1~cm length between adjacent points for the $z$ and $x$ axis, respectively. 

\begin{figure}[h!]
	\begin{center}
		\includegraphics[width=0.48\textwidth]{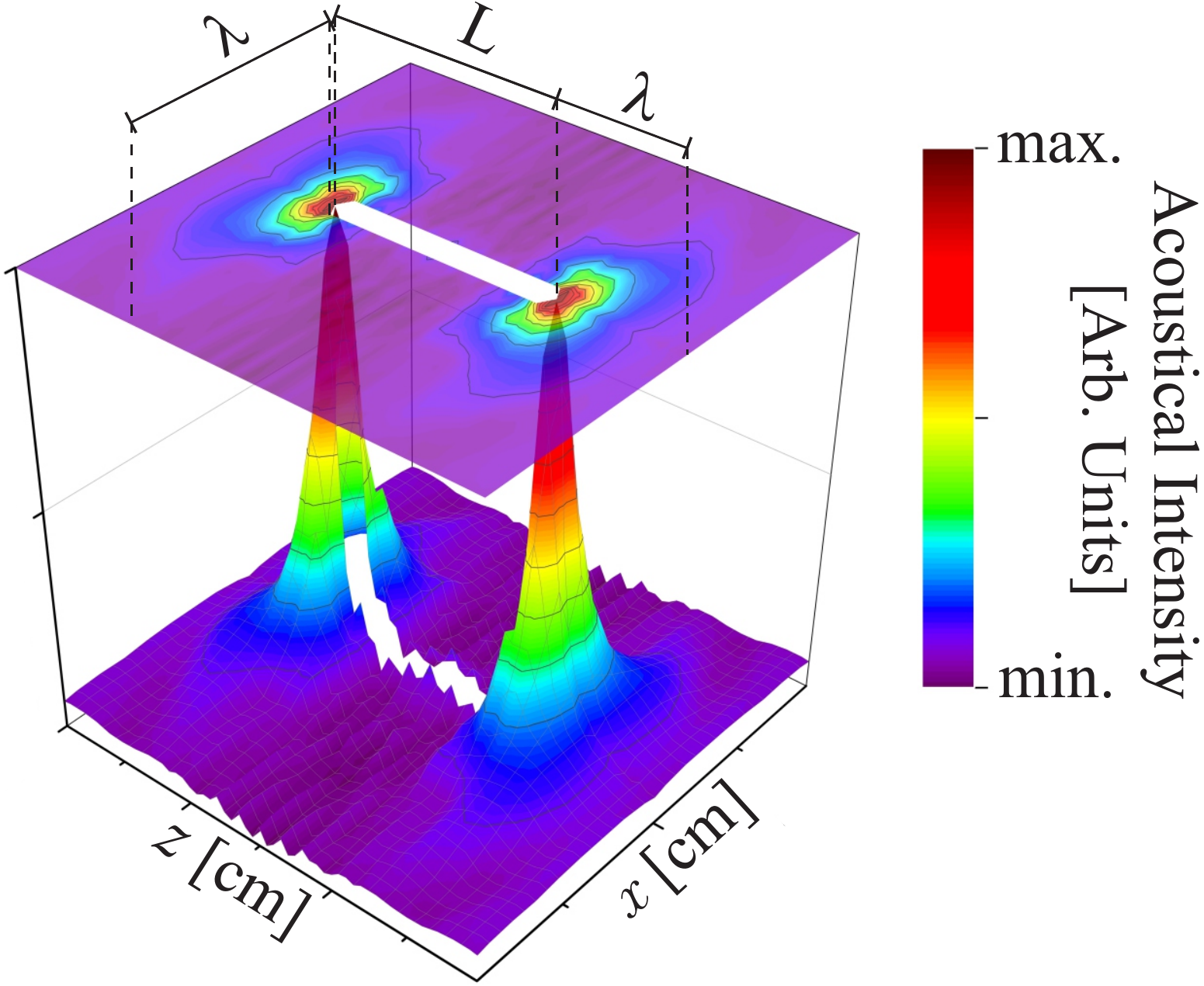}
		\caption{\label{fig:fig4}{(Colour online) The experimental acoustical intensity field radiated by a cylindrical rod vibrating in its first compressional mode. The white rectangle represents the physical space occupied by the rod. As the elasticity theory predicts, the maximum displacement is at the rod edges and the node is placed at the geometrical center of the rod, Eqs.~(\ref{eq:freq}) and (\ref{eq:nodes}). Here we plot the 3D acoustical intensity field as well as a projection in 2D at the top where contour lines are shown.}}
	\end{center}
\end{figure}

From each measurement in regions A and D, the extracted maximum value of the amplitude is assigned to the spatial coordinate where the microphone is placed. On the other hand, each coordinate in regions B and C has two different measurements related to the orientation of the microphone. The actual value for those points is obtained by computing the square root of the components square addition. Fig.~\ref{fig:fig4} shows a 3D plot of the experimental acoustic intensity field as well as its 2D projection. It can be observed the maxima of the field are found at the ends of the rod while the field is weakest near the geometrical center of the rod, as was expected.

\section{Semi-Analytical approach: A simplified model}

\begin{figure}[h!]
	\begin{center}
		\includegraphics[width=0.5\textwidth]{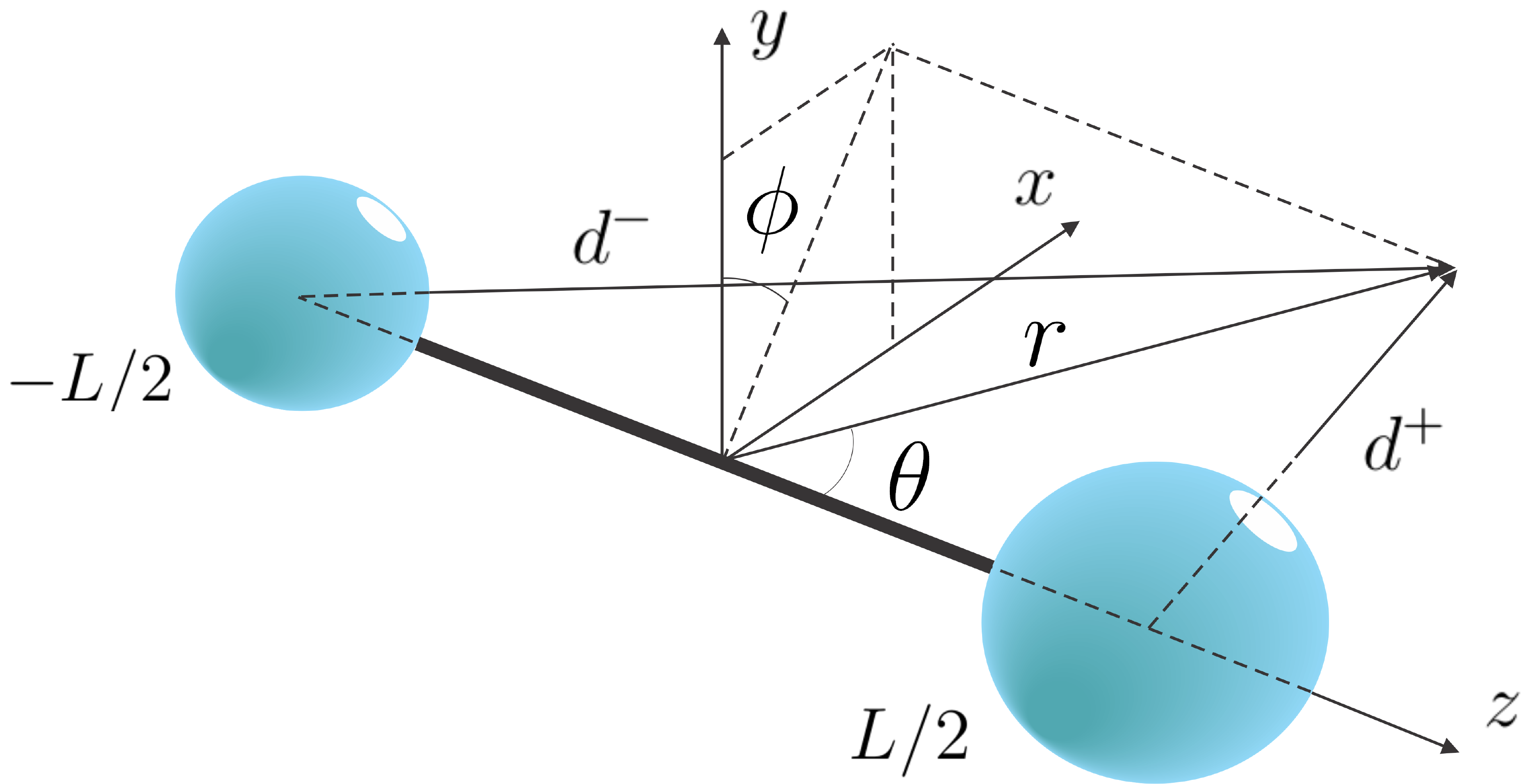}
		\caption{\label{fig:fig5}{(Colour online) Schematic diagram of two identical spheres vibrating in free space separated by a distance $L$. The spurious thick black line that links both spheres does not interfere with the vibration of them but suppresses the acoustic field through all its length.}}
	\end{center}
\end{figure}

In acoustics, spherical sources are good approximations to describe the radiated sound of many structures~\cite{Morse, Kinsler, Rossing2004, Rienstra}. As an approximation to the experimental rod, we place two spherical sound sources separated by a distance $L$. The spheres are joined by a narrow rod which does not interfere with their vibrations and does not emit sound. In other words, we have two spherical sources emitting a superposition of spherical outgoing waves, such that the wavefield along the connecting segment vanishes (see Fig.~\ref{fig:fig5}).

The sound intensity is defined as the average energy flux corresponding to sound propagation. Then, in terms of the complex fields at a point ${\bf{r}}$ it can be written as
\begin{equation}
\label{eq:intensity}
{\bf I}({\bf r})=\frac{1}{2} \textrm{Re}\left[p({\bf r}){\bf u}^*({\bf r})\right],
\end{equation}
where ${\bf u}$ is the acoustic velocity related with the acoustic pressure $p$ through
\begin{equation}
\label{eq:u}
{\bf u}({\bf r})=\frac{-\textrm{i}}{\omega\rho}\nabla p({\bf r}),
\end{equation}
being $\rho$ the density of the medium, and $\omega$ the oscillation frequency. The time-independent sound pressure field is described by the 3D-Helmholtz equation
\begin{equation}
\label{eq:Helmholtz}
(\nabla^2 + k^2)p({\bf r}) = 0,
\end{equation}
where $k$ is the wavenumber and $\nabla^2$ is the Laplacian operator. As might be expected, the suitable coordinate system to describe the pressure field is the spherical one. The frequently encountered definitions where $r$, $\theta$, and $\phi$ give the radial distance, the polar, and the azimuthal angle, respectively, are taken. The coordinate system can be observed in Fig.~\ref{fig:fig5}. 

In addition to the rotational symmetry of the problem, we know the boundary conditions respect to the coordinates, one at the surface of the sphere and another one at infinity where the intensity field must vanish~\cite{Jacobsen}. Thus, the boundary conditions that the acoustic pressure must be fulfilled at the surface are~\cite{Zaman}
\begin{equation}
\label{eq:boundary}
\frac{\partial p(r,\theta)}{\partial n}\bigg\rvert_{S}=\hat{\bf n}\cdot\nabla p(r,\theta)\bigg\rvert_{S}=0,
\end{equation}
with $\hat{\bf n}$ the normal vector to the surface, $S$, of the ``rod''. Also, the imposition of Eq.~(\ref{eq:boundary}) leads to maximum values of the pressure at the surface of the ``built rod''.

In order to find the solution of Eq.~(\ref{eq:Helmholtz}) for a single sphere, it is possible to apply the separation of variables method to describe the pressure field as a product of functions for each one of the coordinates $R(r)$, $\Theta(\theta)$, and $\Phi(\phi)$. Thus, the solution is a product of spherical harmonics and a superposition of outgoing and incoming spherical Hankel functions. However, if a temporal harmonic oscillation of the form $\textrm{e}^{-\textrm{i}\omega t}$ is considered, $h_n^{(1)}(kr)$ can be interpreted as outgoing traveling spherical waves and $h_n^{(2)}(kr)$ as the incoming waves which must be removed. Therefore, the resulting time-independent solution is a sum of the spherical harmonics times the outgoing spherical Hankel functions
\begin{equation}
\label{eq:3Dgensol}
p(r,\theta,\phi) = \sum_{n,m=0}^{\infty} A_{nm} h_n^{(1)}(kr)Y_n^m(\theta,\phi),
\end{equation}
where the coefficients $A_{nm}$ are, in general, complex.

Let us add another sphere separated a distance $L$ and a narrow rod that joins both spheres which does not interfere with their vibrations nor does not emit sound. Thus, the whole pressure field is the superposition of the field generated by each sphere placed at $z=\pm L/2$, therefore
\begin{equation}
\label{eq:ptot}
p(r, \theta, \phi) = p^+(r, \theta, \phi) + p^-(r, \theta, \phi).
\end{equation}
Here, the $\pm$ sign labels the corresponding pressure of the sound emitted by the sphere located at $\pm L/2$, respectively. In general, the pressure field depends on the three cordinates, however we only consider the azimuthal symmetric case $m=0$ due to the studied system. Therefore, Eq.~(\ref{eq:3Dgensol}) for each sphere becomes
\begin{eqnarray}
p^\pm(r,\theta) &=& \sum_{n=0}^{\infty} \sqrt{\frac{2n+1}{2}}A^\pm_{n} h_n^{(1)}(kd^\pm)\nonumber\\
&\times&P_n\left(\frac{r\cos{\theta}\mp L/2}{d^\pm}\right),
\end{eqnarray}
with $d^\pm=\sqrt{r^2\mp rL\cos{\theta}+ L^2/4}$, and $P_n$ being the Legendre polynomial of degree $n$.

\begin{figure}[h!]
	\begin{center}
		\includegraphics[width=0.48\textwidth]{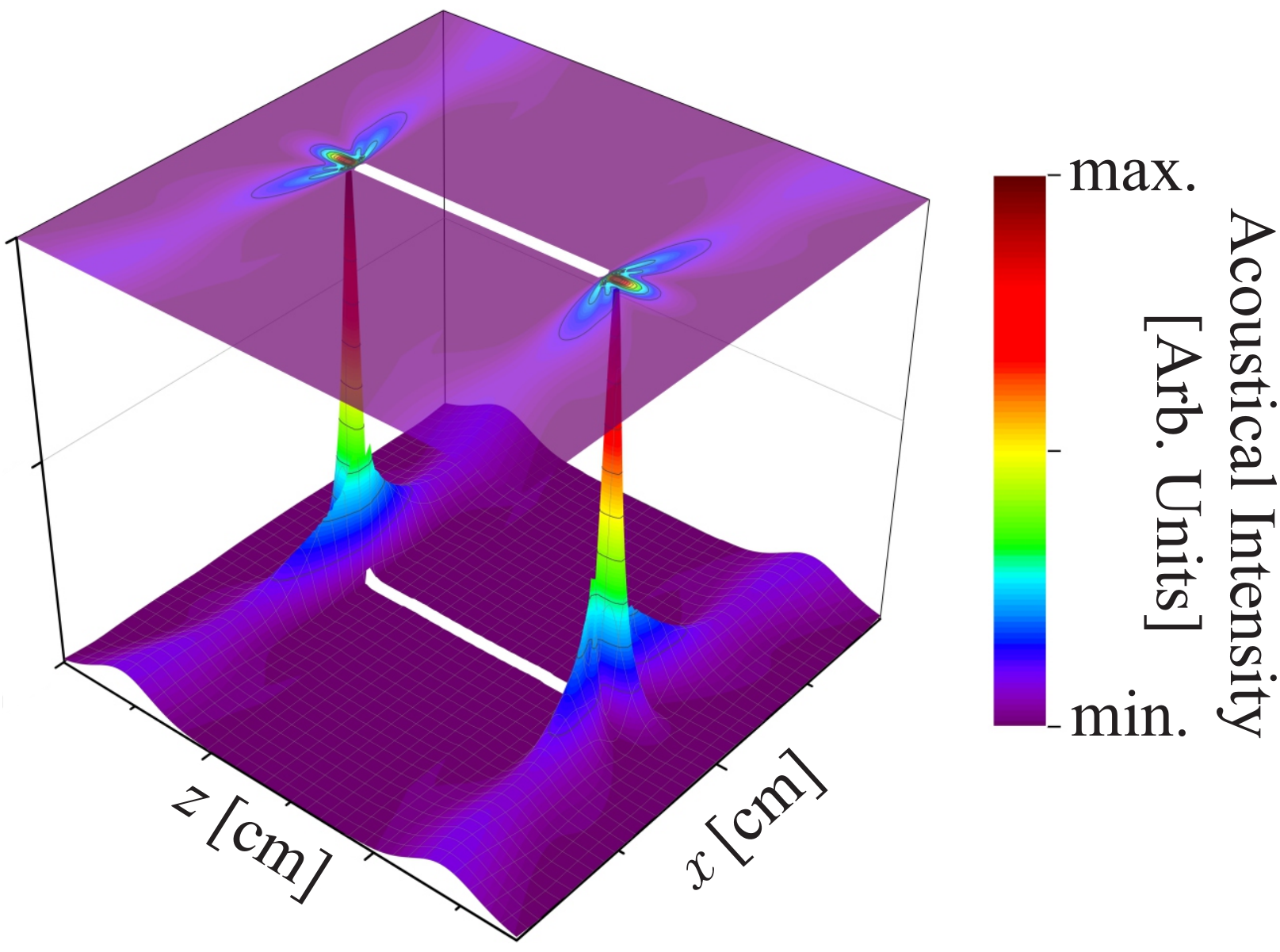}
		\caption{\label{fig:fig6}{(Colour online) The acoustical intensity field radiated by two spherical sources jointed by a narrow beam. The white rectangle represents the physical space occupied by the ``rod''. Here we plot the 3D acoustical intensity field as well as a projection in 2D at the top where contour lines are shown. The numerical parameters used to obtain this graph are $k=0.45$, $L=50$, $A^\pm_0=1$ and $r_0=13$ as well as the normalized units $2\rho\omega=1$.}}
	\end{center}
\end{figure}

Fig.~\ref{fig:fig6} shows the intensity acoustic field $\rvert{\bf{I}}({\bf{r}})\rvert$ radiated by our model considering $n_{\rm{max}}=13$. Due to the divergence of Neumann functions at $kr=0$, we consider an initial radius $r_0$ such that the evaluation can be performed. This value is interpreted as the physical radius of each sphere plus the distance between the surface and the measurement device.

\begin{figure}[h!]
	\begin{center}
		\includegraphics[width=0.48\textwidth]{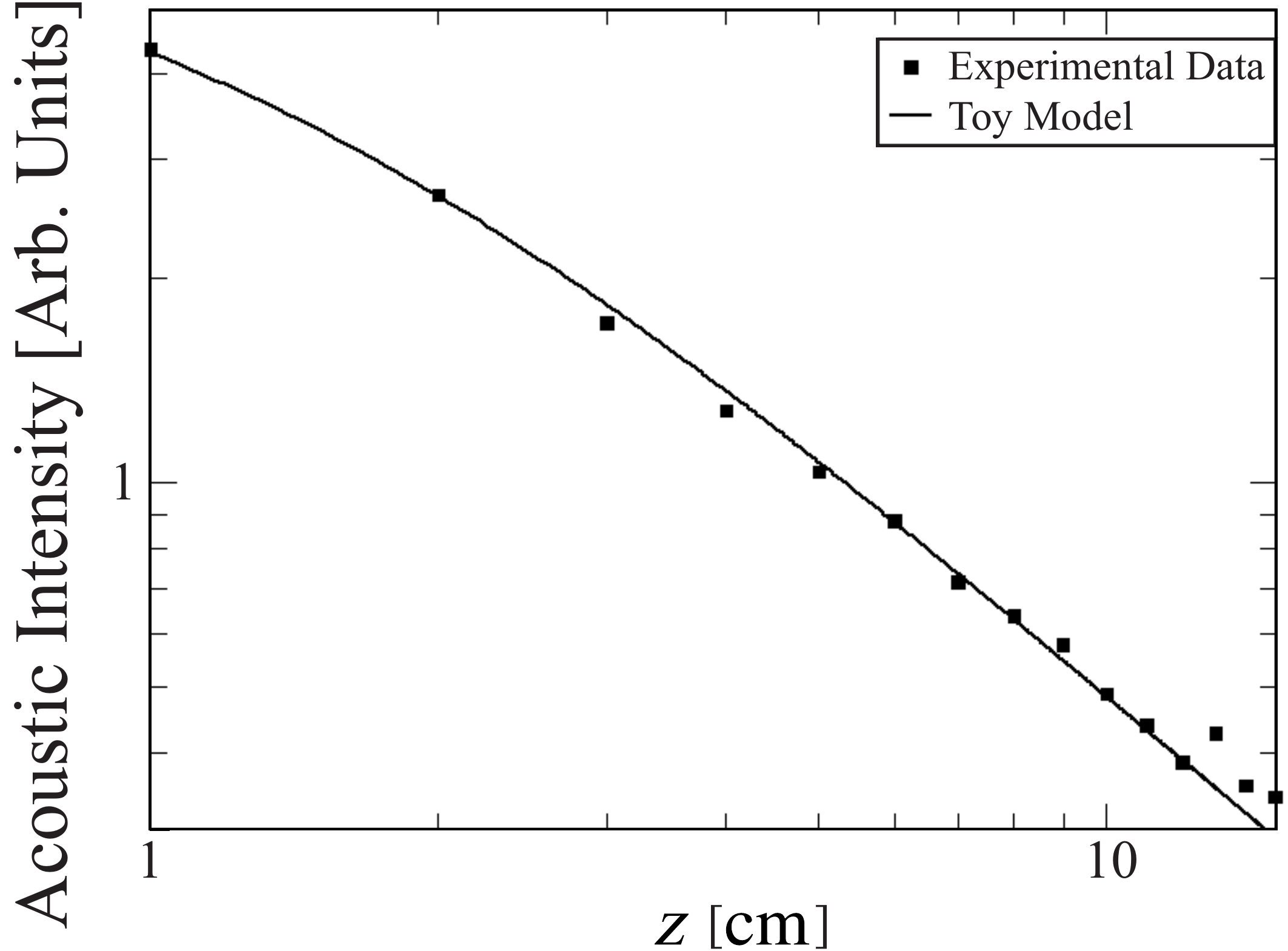}
		\caption{\label{fig:fig7}{Comparison between experimental data of region A (dots) and the fitted absolute value of Eq.~\ref{eq:fitting} (black solid line).}}
	\end{center}
\end{figure}

From Fig.~\ref{fig:fig4} and Fig.~\ref{fig:fig6} the acoustic intensity of both, experimental and semi-analytical, systems can be compared qualitatively. For a deeper analysis, we look for a quantitative comparison of both behaviors. In Fig.~\ref{fig:fig7} the decay of the acoustic intensity field of region A is plotted. It can be observed the experimental data (black squares) and the semi-analytical result (black solid line). It is noticiable that the analytical curve has a good agreement with experimental data.

The semi-analytical curve plotted in Fig.~\ref{fig:fig7} is an evaluation of 
\begin{equation}
\label{eq:fitting}
I(z)=a\textrm{Re}\left[{\rm{i}}p(z+c)\partial_z p^*(z+c)\right],
\end{equation}
by using the amplitudes $A_n$ obtained from Fig.~\ref{fig:fig6}. Here, $a=42.4$ is a general amplitude, $b=0.5$~cm$^{-1}$ the wavenumber and $c=2.925$~cm a traslation in the $z$-axis. By taking $b$ and the resonance frequency as $f_{\rm{C}}=2.5$~kHz, one can compute the speed of sound as ${\rm{c}}\approx314$~m/s. The accuracy of the fitting was found as $R-$square$=0.9969$.

\section{Conclusions}

In this work we have studied experimentally the time-independent acoustic intensity field radiated by a cilindrical rod vibrating in its first compressional mode. The rod was excited by an EMAT together with a VNA, and the acoustical response was measured with a microphone. The resulting acoustical field was obtained by mapping the surrounding area of the rod with the microphone. To avoid undesired wall reflections, a system of foams was developed to mimic an anechoic chamber but at a very low cost.

To model the behavior of the intensity field, we proposed a simple analytical model of two spherical sources joined by a non-interacting thin rod. We added the constriction of no acoustical emission from the rod that joins the sources and the proper boundary conditions. As a result, we obtained a good qualitatively agreement comparison between our model and the experimental measurements.

Finally, we performed a quantitative comparison between the experimental measurements and the analytical model for one of the studied spatial regions with an excellent agreement.

\begin{acknowledgments}
	V.~D.-R. thanks the financial support of DGAPA. The authors thank Centro Internacional de Ciencias A.~C. for the facilities given for several groups meetings and gatherings celebrated there, as well as the given space to set up a laboratory. We also thank M.~Mart\'inez-Mares and G.~B\'aez for usefull comments. L.~A.~R.-L acknowledges support by Academia Mexicana de Ciencias.
\end{acknowledgments}

\end{document}